# Extraordinary magnetoresistance in high-quality graphene devices with daisy chains and Fermi-level pinning

Bowen Zhou,[1,*] Kenji Watanabe,[2] Takashi Taniguchi[3]

[1]*Department of Physics, Technical University of Denmark, 2800 Kgs. Lyngby, Denmark*

[2]*Research Center for Electronic and Optical Materials, National Institute for Materials Science, 1-1 Namiki, Tsukuba 305-0044, Japan*

[3]*Research Center for Materials Nanoarchitectonics, National Institute for Materials Science, 1-1 Namiki, Tsukuba 305-0044, Japan*

[*]To whom all correspondence should be addressed: nerozh@outlook.com

**Abstract**

We studied daisy-chained extraordinary magnetoresistance (EMR) devices based on high-quality monolayer graphene encapsulated in hexagonal boron nitride (h-BN) at room temperature. The largest magnetoresistance (MR) achieved in our devices is $4.6 \times 10^7$ %, the record for EMR devices to date. The magnetic field sensitivity, *dR/dB*, reaches 104 kΩ/T, exceeding the previous record set by encapsulated graphene by more than 300 %, and is comparable with state-of-the-art graphene Hall sensors at cryogenic temperatures (4.2 K). We demonstrate that daisy chaining multiple EMR devices is a new way to reach arbitrarily high sensitivity and signal-to-noise ratio, and extremely small noise equivalent field for weak magnetic field detection. Finally, we show the evidence of metal contact-induced Fermi-level pinning in the sample and its influence on graphene properties, current distribution and EMR performance. We highlight the EMR geometry as an interesting alternative to the Hall geometry for fundamental physics studies.

**Main**

The advent of giant magnetoresistance (GMR) [1] and colossal magnetoresistance (CMR) [2] has led to significant scientific and technological advancements in magnetic sensing devices and hard disk drives, and has inspired the search for other magnetoresistance effects. In 2000, the extraordinary magnetoresistance (EMR) effect was discovered by Solin et al. [3]. EMR is characterized by a change in the current flow owing to the Lorentz force without the use of magnetic materials. Although the EMR effect shares similarities with the Hall effect, it employs a hybrid architecture with carefully shaped regions of different conductance to further enhance the magnetoresistance,

$$MR = \frac{R(B)-R_0}{R_0}. \quad (1)$$

Here, R(B) represents the resistance of the device at magnetic field B and $R_0$ the resistance at zero magnetic field. Note, $R_0$ can be replaced by $R_{min}$ if the minimal resistance is reached at a finite positive or negative magnetic field [4]. Typically, EMR is enhanced when there is a significant difference in conductivity between the high- and low-conductivity regions as well as a high carrier mobility in the low-conductivity region [5], where the high- and low-conductivity materials could be metals and moderately doped semiconductors, respectively. Notably, a magnetoresistance of approximately $10^7$ % was recently achieved in encapsulated graphene EMR devices [6] at T = 300 K and B = 9 T. This MR surpasses that of any other reported device and is one order of magnitude greater than the values found for bulk 3D semiconductor devices [3].

Most previous research on EMR magnetometry sensors focused on maximizing their MR [3,5,7,8]. In devices and materials with exceptionally high carrier mobility, such as III-V materials or encapsulated graphene, the elastic mean free path may surpass the size of the device [6,9,10]. In this ballistic transport regime, $R_0$ can approach zero [6] and result in an extremely large and unsaturated MR value, $4.6 \times 10^7$% (Figure 1(d)), the record for EMR devices to date. In this regime, $R_0$ can also change sign [6] and lead to negative MR values that may not necessarily translate into practical use. For magnetic sensing, the magnetic field sensitivity *dR/dB* serves as a more appropriate figure of merit because it captures the change in resistance in response to a magnetic field and does not depend on $R_0$. More importantly, chasing a large MR record may not lead to many practical applications, however, competing for a small magnetic field detection limit does as the cheap and convenient room-temperature magnetic field sensors are desired for many applications such as magnetic navigation [11], electromagnetic non-destructive testing [12], magnetic based medical diagnosis to detect weak magnetic fields generated by neurons and map brain activity [13–15]. To reach a small detection limit for EMR sensor, a high sensitivity is needed (Equation (3) and (5)).

This study evaluates the feasibility of achieving high sensitivity for EMR in magnetic fields. To this end, we fabricated microscale EMR devices based on graphene encapsulated within hexagonal boron nitride (hBN) flakes and experimentally measured a room temperature sensitivity as high as 104 kΩ/T at a magnetic field strength of 0.2 T (experiment details in Methods). This finding is comparable to the highest sensitivity reported for a graphene Hall sensor measured at a temperature of 4.2 K [16]. All measurements in this study were performed at room temperature.

We demonstrate that a decrease in the charge carrier density and daisy-chaining of multiple EMR devices can enhance the sensitivity, and then discuss the potential contribution of ballistic effects to further boost the performance. Finally, we consider the possible role of metal-graphene contact-induced Fermi-level pinning on graphene properties and EMR behavior. In the figures of this manuscript, we use R and dR/dB to represent 4-terminal resistance and sensitivity, 2-pt R and 2-pt dR/dB to represent 2-terminal resistance and sensitivity.

**Experiment**

**1. High record of magnetoresistance and charge neutrality points**

Figure 1(a) shows the room-temperature zero-field four-terminal resistance of the three devices, with the current flowing through all three devices from terminals 3 to 9 (the leftmost probe to the rightmost probe) via the two large connecting electrodes. All three samples were electron-doped with charge neutrality points (CNPs) close to − 9 V. The asymmetry in zero-field resistance vs. $V_g$ can be ascribed to the band re-alignment and a resultant pn-junction near the metal-graphene interface [6,17][44]. Figure 1(b) shows the resistance of the small device R(B) with a radius of $r_0 = 2.6\ \mu m$ in response to a magnetic field B at different gate voltages $V_g$ from 4-terminal measurements. R(B) has a maximum of ~ 12 kΩ at -9T, close to the CNP, and drops off as $V_g$ is tuned away from the CNP. At $V_g$ = -4 V, R(B) decreases to 4 kΩ at -9 T.

The largest MR value we found at room temperature in this set of EMR devices is more than $4.6 \times 10^7$% and unsaturated to 9 T in the small device at $V_g$ = -7 V (Figure 1(d)) corresponding to the light blue resistance trace in Figure 1(b). The corresponding minimal resistance $R_{min}$ is as small as 0.02 Ω that could be ascribed to room temperature ballistic transport [6,9,10], and can be accurately detected by a long measurement time to average out noise and increase signal-noise ratio.

Figure 1(c) shows the 4-terminal sensitivities dR/dB of the small device in response to the magnetic field at different $V_g$ values calculated from the slope of the resistance, suggesting that the sensitivity is also the largest (2 kΩ/T) near the CNP and decreases when $V_g$ is tuned away from it. The sensitivity at +/-1T for different $V_g$ values is shown in Figure 6, confirming that the maximum is reached near the CNP. Similarly, the magnetic field sensitivity of the Hall sensor, also known as the Hall coefficient, is very large at low carrier densitities, and also increases with increasing mobility [18,19]. In addition, the conduction in high-quality graphene devices is limited by electron-phonon scattering rather than charged impurity scattering, so graphene tends to exhibit higher mobility near the CNP [20] which also improves the sensitivity of EMR devices [5]. Therefore, to maximize the magnetic field sensitivity of the EMR device, a low charge density in graphene or a low charge density bulk material with a high mobility can be considered.

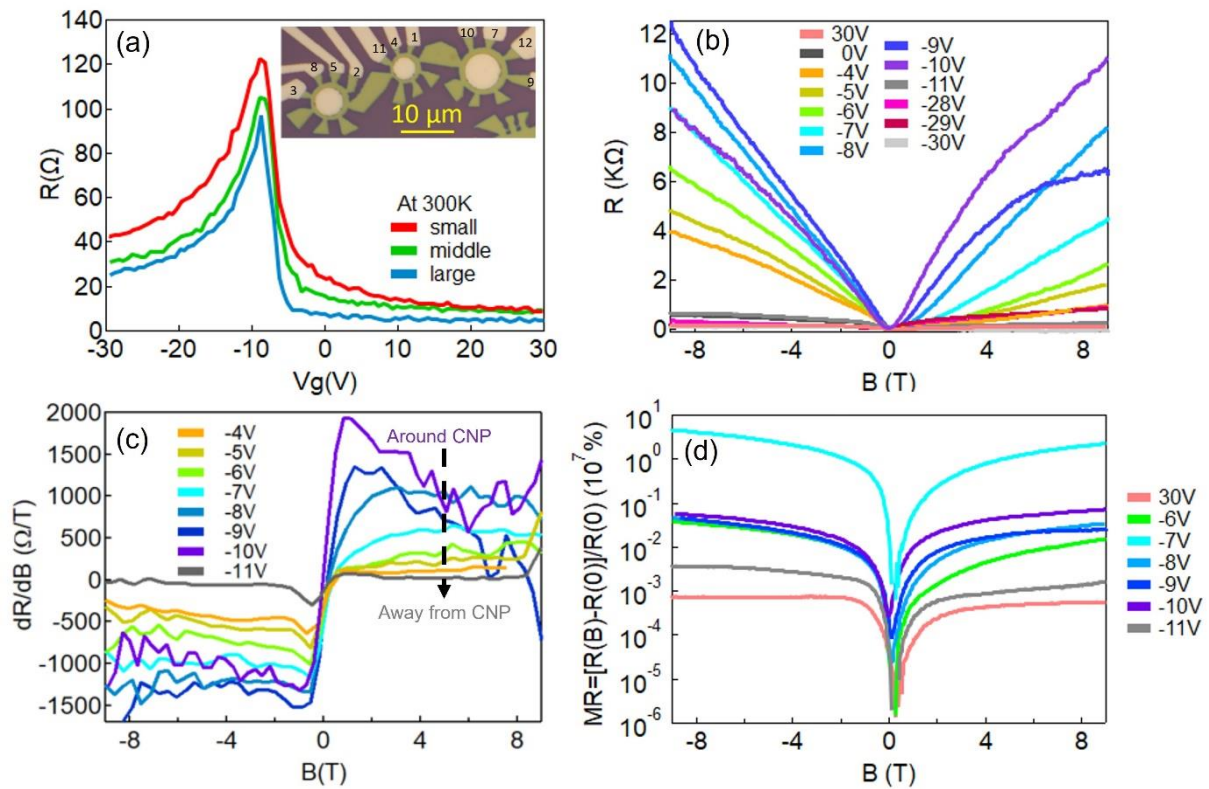

*Figure 1. (a) Zero-field resistance as a function of gate voltage with CNP at -9V. The inset shows the optical image of three devices. (b) Resistance R as a function of magnetic field B at different gate voltages Vg from 4-terminal measurements. The near-zero-resistance range version is shown in Figure 7. (c) Sensitivity dR/dB as a function of magnetic field B at different Vg values calculated from (b), suggesting that the sensitivity is the largest near the CNP and decreases away from it. (d) Magnetoresistance MR as a function of magnetic field B at different gate voltages Vg from 4-terminal measurements with the largest MR value measured to date in EMR devices. Note the log-scale of the MR axis.*

## 2. The highest room temperature sensitivity

As previously reported [6,21], the 2-terminal sensitivity can be more than one order of magnitude larger than the 4-terminal sensitivity for the same EMR device. Figure 2(a) shows the room-temperature serial resistance of all three devices as a function of $B$ and $V_g$, measured in the 2-terminal configuration at terminals 3 and 9. Similar to the 4-terminal measurements (Figure 1), the 2-terminal resistance at the CNP exhibited a pronounced enhancement in response to the magnetic field. Figure 2(b) shows the corresponding 2-terminal sensitivity $(dR/dB)_{2T}$ is indeed maximal near the CNP, but also with a tendency to be highest around $B = -0.2$ T. Figure 2(c) and 2(d) show the extracted 2-terminal R and dR/dB in response to B, respectively. We find that the 2-terminal sensitivity of the devices at room temperature reaches 104 kΩ/T near the critical point B (-0.2 T) and close to the CNP. This value surpasses the previous record set by encapsulated graphene [6] by more than 300 %. The largest sensitivity found in other serially connected EMR devices is 70 kΩ/T (see Figure 8). In contrast, a room temperature sensitivity of 5.7 kΩ/T was reported for an encapsulated graphene Hall sensor [22]. The highest room temperature sensitivity achieved in our EMR devices is comparable to the state-of-the-art graphene Hall sensor at 4.2 K [16], suggesting that graphene EMR sensors could make high-performance magnetometry available in practical and affordable room temperature systems. When B > 0 T, the 2-terminal sensitivity of our devices reaches 58 kΩ/T at 0.2 T near the CNP.

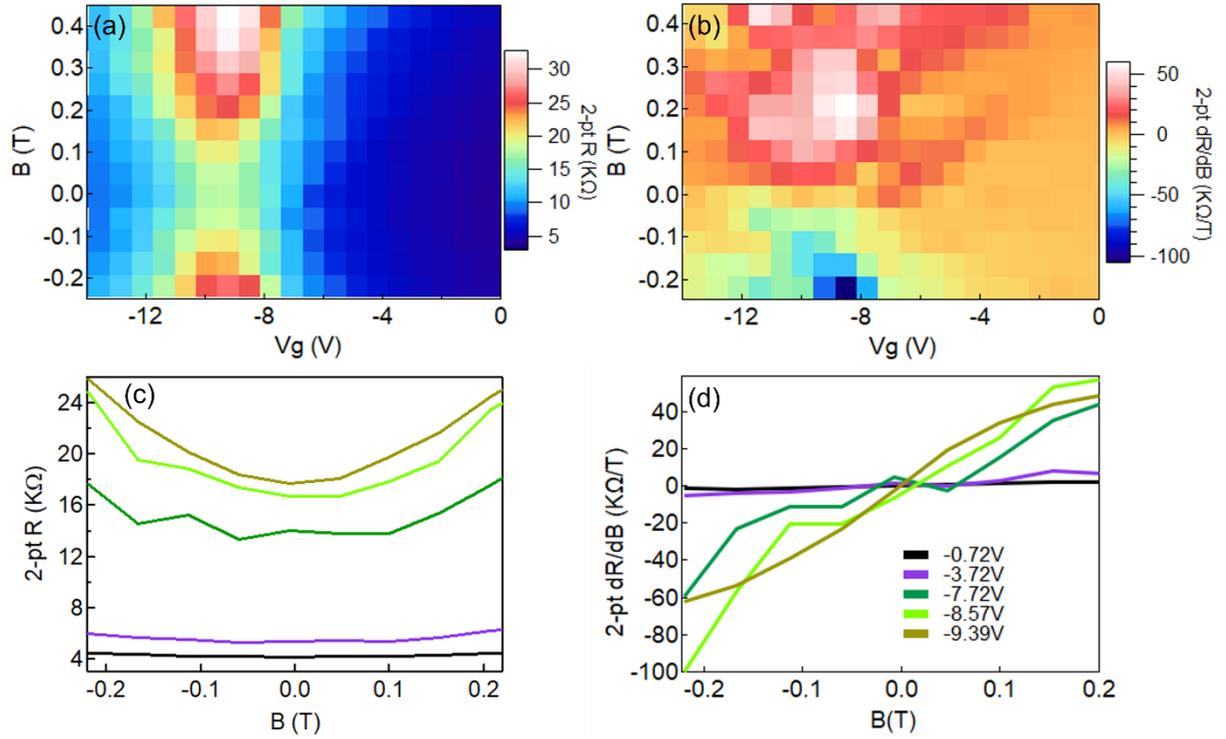

*Figure 2. (a) Room-temperature resistance as a function of B and Vg from two-terminal measurements. (b) Room-temperature two-terminal sensitivity as a function of B and $V_g$ based on (a). (c) Two-terminal resistance and (d) two-terminal sensitivity as a function of B at different Vg are extracted from (a) and (b), respectively. (c) and (d) share the same legend.*

**Simulations**

1. **Arbitrarily high sensitivity by daisy chaining devices**

Since the 2-terminal sensitivity of three-device configuration is significantly higher than that of a single device (Appendix G), we consider the possibility of enhancement of sensitivity when daisy-chaining three EMR devices. In the simplest case, the resistance of three serially connected identical EMR devices is simply the sum of the individual resistances, $R_{total} = 3R$ as the current distribution in each device is identical (insets of Figure 3(b)). The higher sensitivity for *N* identical EMR devices connected in series follows trivially from $dR_N/dB = d(NR_1)/dB = NdR/dB$, where $R_N$ is the serial resistance of all N devices. The magnetoresistance, however, is not increased.

To verify the assumption that devices can be daisy-chained for higher sensitivity, we carried out finite element calcuations on serially connected EMR devices, as shown in Figure 3 (model details in Methods). Ten devices connected in series are shown in Figure 3(a) with the device numbers corresponding to the numbers in the legend of Figure 3(c). Figure 3(b) and (c) show the resistance of each individual device (1 to 5) (also in Figure 9), the total resistances for various combinations of devices 2 to 5 (purple, magenta, grey) as well as all devices in series (red), and their corresponding sensitivities, respectively. There is a consistent incremental increase in sensitivity with the addition of each device in series, as depicted in Figure 3(c), the same as resistance (Figure 3(b)). The total sensitivity of the 10 serially connected devices (red) is almost identical to the sum of sensitivity (black) defined as the sensitivity of device 5 (dark blue) multiplied by 10, except for a small offset attributed to the slightly different geometry of devices 1 and 10 on the two ends. This observation leads to a straightforward extension: for N devices in series, the sensitivity scales linearly as N times the sensitivity of a single device, expressed as $dR'/dB = d(NR)/dB = NdR/dB$. For example, 1000 times enhancement in sensitivity could be reached in 1000 daisy-chained devices. Note that a larger input voltage may be required for the large-number daisy-chained devices in a constant voltage measurement. More experimental tests are needed before practical applications.

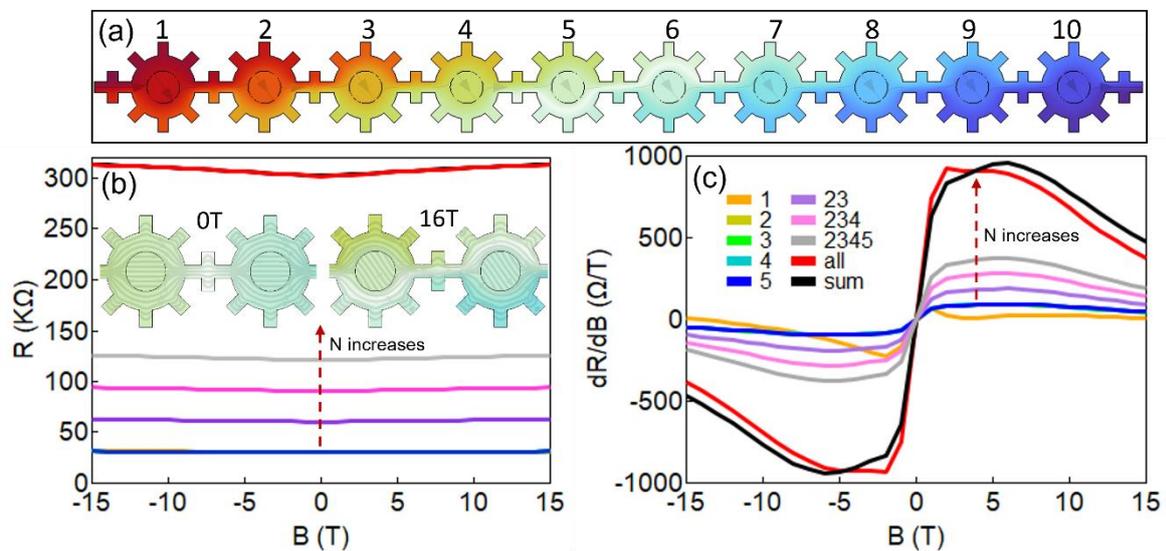

Figure 3. (a) Finite element simulations of 10 EMR devices in series with two vertical rectangular bars on two sides of each device as voltage probes; the color represents the calculated voltage potential with red for positive, blue for zero, and light green for medium. (b) and (c) share the same legend and show the calculated resistances for each device (1 to 5) and various combinations of devices 2 to 5, as well as all devices as a function of magnetic field, and the corresponding sensitivities, respectively. The numbers in the legend denote the device numbers shown in (a), for example, 1 represents device 1, 2345 represents devices 2-5 in series. The resistance and sensitivity of each device (2 to 5) overlap because the devices are identical. The black and red curves designate the sum of the individual sensitivities defined as the sensitivity of device 5 multiplied by 10, and the total (combined) sensitivity of all devices, respectively. The sensitivity of device 1 (orange) is different from the other devices (2-5) due to the asymmetric geometry of device 1 that alters the current distribution, which contributes to the small offset between the sum of individual sensitivities and the total sensitivity. The two insets in (b) show current distribution in device 5 and 6 at zero field and 16T, respectively.

The ability of EMR devices to operate in a 2-terminal configuration makes daisy-chaining EMR devices a straightforward way to reach an arbitrarily high sensitivity. While chaining of multiple Hall bars has been demonstrated, the complicated connection and fabrication processes can lead to many problems and inhibit its applications [23–27]. Our 2-terminal connected EMR devices, however, allow the daisy chaining of devices to be simplified greatly, as all connections can be made on-chip in a single layer [28]. In a similar geometry-- a strip pattern geometry, the graphene magnetic field sensing device shows increase of both magnetoresistance and sensitivity with increasing the number of modules [28].

The signal-to-noise ratio (SNR) and noise equivalent field (NEF) of EMR sensor are

$$SNR = \frac{I_{in}\left|\frac{dR}{dB}\right|\Delta B}{\left[\left(\frac{V}{L}\right)^2 \gamma \mu e R \frac{\Delta f}{f} + 4kTR\Delta f\right]^{\frac{1}{2}}}, \quad (2)$$

$$NEF = \frac{\left[\frac{\left(\frac{V}{L}\right)^2 \gamma \mu e R}{f} + 4kTR\right]^{\frac{1}{2}}}{I_{in}\left|\frac{dR}{dB}\right|}, \quad (3)$$

where $I_{in}$ is the input current, V is input voltage, L is the spacing of the current (voltage) leads, γ is the dimensionless Hooge parameter, e is the electron charge, f is the operating frequency, Δf is the detection bandwidth, k is Boltzmann's constant, T is temperature in Kelvin, R is the two-terminal resistance of the EMR devices [29–31]. The first term in the denominator of Equation (2) is the $1/f$ noise while the second term is the thermal noise. We can use a high operating frequency, for example, f=1kHz to reduce the $1/f$ noise [16]. Then the signal-to-noise ratio and the noise equivalent field can be approximated as

$$SNR = \frac{I_{in}\left|\frac{dR}{dB}\right|\Delta B}{[4kTR\Delta f]^{\frac{1}{2}}}, \quad (4)$$

$$NEF = \frac{[4kTR]^{\frac{1}{2}}}{I_{in}\left|\frac{dR}{dB}\right|}. \quad (5)$$

The other advantages of daisy chaining EMR sensors are that SNR increases by $\sqrt{N}$ and NEF decreases by $\frac{1}{\sqrt{N}}$ with the increasing number N of the EMR devices with a constant input current $I_{in}$; here, R is the two-terminal resistance of the EMR devices that follows trivially from the serial resistance of all devices connected in series $R_N = NR_1$. But when N is very large, a higher $I_{in}$ could be used to help increase SNR and reduce NEF as suggested in ref [16]. Such a chain of EMR devices could be meander-shaped, easily incorporating $10^2$-$10^3$ devices, in a region of $0.5 \times 0.5\ mm^2$, and reaches a correspondingly high levels of sensitivity and signal-to-noise ratio, as well as a small noise equivalent field.

## 2. Fermi level pinning

Compared to conventional MR traces of semiconductor devices that tend to be parabolic below medium field [3], the resistance traces of encapsulated graphene EMR devices generally exhibit additional features and each seems to be the combination of two traces with different slopes, as clear from the experimental data for $V_g = -11$ V (black solid line) in Figure 4(c-d) and Figure 10, corresponding to Figure 1(b). Similar features also appear in Figure 7.

We previously proposed that Fermi level pinning (FLP) [32–36] can alter the conductivity of graphene near the metal by shifting the position of graphene Dirac cone in relative to the Fermi level [6] (Figure 11). In the following we examine whether this is a plausible explanation for the observed features using finite element simulation in COMSOL (model details in Methods).

To represent FLP, we use the simplest possible model for introducing a scenario with two different conductivity regions. Instead of a relaxation function [6], a uniform conductivity with an average value is assumed within a region surrounding the metal shunt (Figure 4(b)). As seen in inset of Figure 4(b), we keep the width of the FLP region fixed at $w_{\text{FLP}} = 0.5\ \mu m$, i.e., $r_{\text{FLP}} = r_{\text{Shunt}} + w_{\text{FLP}}$, where $r_{\text{Shunt}}$ and $r_{\text{FLP}}$ are the radii for the shunt and shunt+FLP region, respectively. The outer region is non FLP region of graphene. The fitting with experiment data was carried out manually by inspecting of the similarity after extensive parameter sweeps of conductivity and carrier mobility. We give here two sets of values for $(\sigma, \mu)$ for both regions that represent the best agreement with the data we were able to achieve: For the outer region, we obtained $(3.48 \times 10^6$ S/m, 10000 cm$^2$/Vs$)$ and $(3.66 \times 10^6$ S/m, 8500 cm$^2$/Vs$)$, while for the inner FLP region we found $(1.74 \times 10^7$ S/m, 30000 cm$^2$/Vs$)$ and $(2.44 \times 10^7$ S/m, 30000 cm$^2$/Vs$)$. In the Figure 4(c) they are denoted "optimal1" (blue solid line) and "optimal2" (red solid line). As expected, both the conductivity and mobility in the inner FLP region are much higher than those in the outer non FLP region. Other representative fits with different carrier mobility are shown in Figure 10. Without the FLP region, the agreement with experimental data was significantly worse, as seen in the dash-dot blue and red curves in Figure 4(c).

We measured the conductivity and carrier mobility from a nearby Hall bar made from the same encapsulated graphene stack as the EMR samples, obtaining the values $\sigma = 5.8 \times 10^6$ S/m and $\mu = 82583$ cm$^2$/Vs. Using these values for simulation of outer region of graphene of the EMR devices and the best values of inner region graphene $\sigma = 6.68 \times 10^6$ S/m and $\mu = 82583$ cm$^2$/Vs (found after parameter sweeps), the green dashed curve was obtained, which shows even less agreement with the experimental data.

In Figure 4(d), the size of the FLP region is changed systematically. With the parameters corresponding to the "optimal2" fit in Figure 4(c), we varied the radius of the FLP region $r_{\text{FLP}}$ from 1.75 $\mu m$ to 2.65 $\mu m$ as shown in the inset of Figure 4(b). The radii 1.75 $\mu m$ to 2.65 $\mu m$ correspond to nearly no FLP region (red line) and FLP extending nearly to the outer left edge of graphehe (grey line), respectively. As expected, the trace for 1.75 $\mu m$ FLP region is equivalent to that without FLP shown in red dash-dot line in Figure 4(c). The radii that lead to the best fits for either positive or negative values of the magnetic fields, are in the range $2 - 2.25$ $\mu m$.

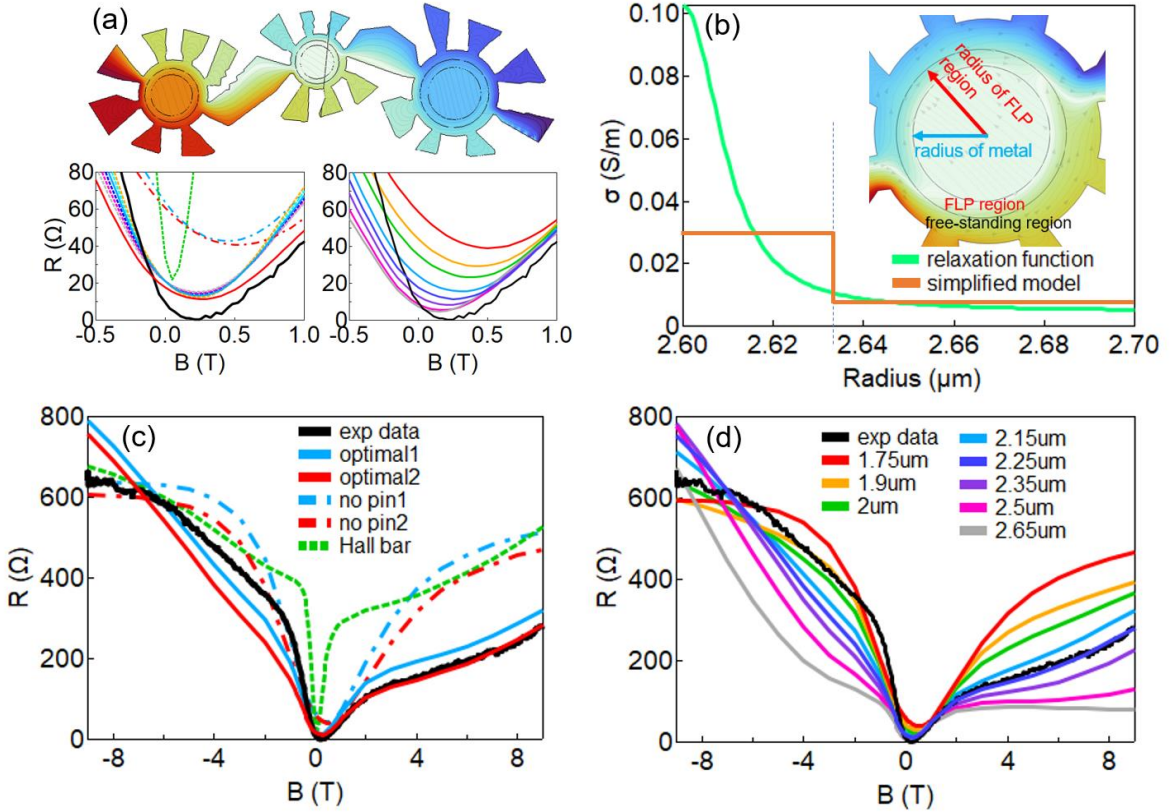

*Figure 4. (a) COMSOL simulation of the three EMR devices shown in inset of Figure 1(a), at B= -9T. The color represents the voltage potential with red for positive, blue for zero, and light green for medium. The two insets are near-zero-field range of (c) and Figure 10, and (d), respectively. (b) Schematics of graphene conductivity varying spatially near the metal-graphene interface. In the model with a relaxation function based on ref [6], the conductivity continuously changes away from metal, and in the bulk, it is free-standing and determined primarily by the applied gate voltage. In our simplified model,*

*the conductivity near the metal and in the bulk are constant. The inset shows geometry of the central device with metal shunt, FLP region and non-FLP graphene regions, respectively. (c) Experiment (black solid line) and simulations of resistance as a function of B. The blue and red solid lines are from two sets of optimal parameters, termed "optimal1" and "optimal2". The two dash-dot lines are optimal simulations without two segmented regions (i.e., without FLP region). The green dash line is based on parameters measured from Hall bar. (d) Simulations of resistance as a function of B with varying radius of FLP region.*

We note that all simulations fail to capture the small resistance near zero field (insets of Figure 4(a)). The measured $R_{min}$ is 0.41 Ω, smaller than resistance of conventional graphene and metal; this could be ascribed to room temperature ballistic transport [6,9,10], which is not supported in finite-element simulations (details in methods). At around 50 mT, the cyclotron radius is comparable to the device dimensions, which suggests that for smaller fields geometric effects might be more important, and may explain why the FEM simulations fits the experimental data best for larger fields (right side of red solid line in Figure 4(c)).

While FLP is typically not observed in bulk semiconductor EMR devices, two reports show that unencapsulated monolayer graphene EMR devices with top-contact metal shunt also do not seem to show effects of FLP [7,37]. This could be contributed by two factors: unencapsulated graphene has a low quality and shows broader resistance traces, thus the effect of FLP may be not obvious; compared to edge-contact graphene, surface-contacted graphene has less overlap of the metal and carbon orbitals and less surface bonding sites, and could have a weak metal-graphene coupling and thus a weak Fermi-level pinning [6,20,32–34,36,38,39][44].

We are not aware of any reported experimental study on FLP in graphene Hall devices. While Hall devices generally have large open areas where the current can run far from edges and contacts, the current in a circular EMR geometry flows adjacent to the metal shunt, which in our case can have a diameter close to the metal shunt (Figure 5). At B = ±9T in COMSOL simulations, a significant portion of current runs through the outer region of graphene where FLP should be less important, while at B = ±2T, a significant portion of current runs through the inner region of graphene where the charge carriers are more affected by the effect of FLP. This supports the picture of the FLP to primarily appear at low field and the two kinks existing in the resistance trace at around ±2T (black solid line in Figure 4(c) and (d)).

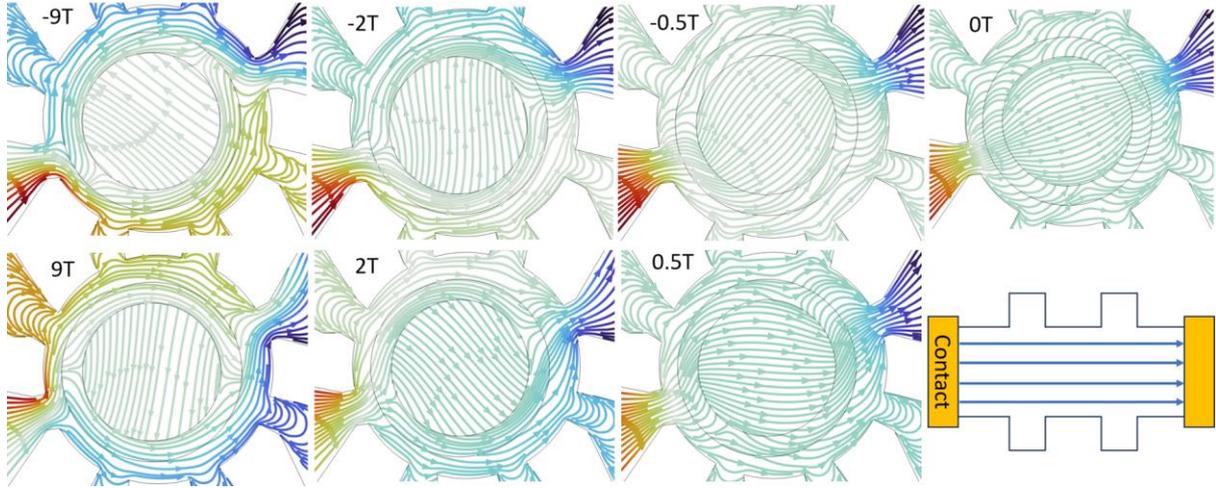

*Figure 5. COMSOL simulation of the central EMR device at B=±9T, ±2T, ±0.5T, and 0T, and schematics of current trajectories in a Hall bar at zero field. The metal shunt, FLP region and free-standing region of graphene are in the inner, middle and outer layers, respectively. The color in simulation represents the voltage potential with red for positive, blue for zero, and green for medium. The current trajectories and densities are represented by arrows and lines.*

We note that the simulations that match the data best, represent significantly lower carrier mobilities for the EMR device, compared to the adjacent Hall-bar. While one explanation could be that the model is too simplified, it is also possible that the graphene in the device really has a lower mobility. As the simulation provides a compelling fit to the data it is worth considering what could lead to a decreased mobility of the graphene. One such possibility is that stress in the e-beam evaporated aluminium (Al) film [40,41] could transfer from the large central Al shunt to the surrounding graphene via the direct edge contact, and induce lattice distortion and strain disorder in the graphene, which similar to corrugations could decrease the mobility and conductivity of graphene in EMR devices [41,42], compared to the adjacent Hall bar. Additionally, different fabrication processes in EMR device and Hall bar could lead to different properties of local samples. The local charge inhomogeneities in the same stack, such as air bubbles trapped between graphene and boron nitride layers, could also lead to different properties.

**Conclusion**

We found the largest magnetoresistance (MR) in our devices, $4.6 \times 10^7$ %, a new record for EMR devices. The potential applications of large magnetoresistance could be magnetic memories, magnetic valves, magnetic sensors and magnetic switches. In terms of magnetometry, the magnetic field sensitivity dR/dB could be a better figure of merit for magnetic sensing compared to MR as it captures the change of resistance with magnetic field and does not depend on the zero-field resistance. We find that the sensitivity can be increased by reducing the charge carrier density in materials, and chaining EMR devices. Daisy chaining EMR devices presents a straightforward way to improve the sensitivity by N while the noise equivalent field decreases by 1/sqrt(N) for weak field detection. The largest room-temperature sensitivity of our EMR sensor is comparable with graphene Hall sensors at 4.2 K, which makes the graphene EMR sensor an interesting alternative to cooled Hall sensors for affordable, high performance magnetometry. The impact of metal-graphene contact-induced Fermi-level pinning on graphene properties and EMR behaviour is also very significant, suggesting EMR geometry as an interesting alternative to the Hall geometry for fundamental physics studies. A graphene EMR sensor could challenge Hall sensor in many applications, including highly sensitive room-temperature and low-temperature magnetic-field sensors, and high-field temperature sensors in cryogenics.

**Methods**

**1. Experiment details**

The devices were manufactured from mechanically exfoliated monolayer graphene encapsulated in 30 nm thick hBN flakes using a dry-transfer technique [20]. The hBN-graphene-hBN stacks were placed on highly p-doped Si substrates with a 300 nm thick $SiO_2$ layer on top. The device geometry was defined by electron beam lithography (EBL) followed by reactive ion etching (RIE), resulting in disk-shaped devices with outer radius $r_o$ and concentric circular holes with inner radius $r_i$. Electrical contacts and a central metal shunt were then defined by EBL and electron beam evaporation of a 4/80 nm Ti/Al electrode layer. The inset of Figure 1(a) shows the optical image of three EMR devices in the series, where the ratio $r_i/r_o$ is constant but $r_o$ is 3.57 um, 2.6 um and 4.46 um, from left to right. The inset of Figure 12(a) shows three EMR devices in series with the same outer radius $r_o$ = 2.6 um, but different inner radii $r_i$ = 1.7 um, 1 um and 1.3 um, from left to right. Electronic transport measurements were carried out using lock-in technique with a 200 μV bias at a frequency of 13 Hz in a variable temperature physical property measurement system (PPMS) equipped with a 9 Tesla magnet. The graphene carrier density $n$ was controlled by a gate voltage $V_g$ applied to the Si substrate. The rate of change in carrier density with gate voltage, $n = V_g \cdot 6.7 \times 10^{10}$ $V^{-1} cm^{-2}$, was determined by a Hall device (without the central shunt) located next to the EMR devices, as the geometry of EMR devices makes it difficult to determine *n* as well as mobility *μ*.

## 2. Simulation details for daisy chaining

We use the commonly-used metal gold in this model based on ref [3,5] to show more generic results. Each EMR device has an outer diameter of 6.02 um and an inner diameter of 3.05 um. The mobility, carrier density and conductivity (μ;n;σ) of the graphene and gold shunt are (0.7 m²/Vs ; $1.03 \times 10^{25}$ m$^{-3}$; $1.15 \times 10^6$ S/m) and (0.00478 m²/Vs ; $1.59 \times 10^{31}$ m$^{-3}$; $1.21 \times 10^{10}$ S/m), respectively. The charge density and conductivity for gold are adjusted to match the thickness of the graphene in the 2D model. The input current density on the leftmost terminal is $1.2 \times 10^8$ A/m², and the rightmost terminal is ground. Two short bars on two sides of each EMR device act as voltage probes. Given that devices 2 to 5 are identical, their resistances and sensitivities are also indentical (Figure 3(b) and (c)). The sensitivity of devices 2, 3, 4 and 5 (grey), that of devices 2, 3 and 4 (magenta), and that of devices 2 and 3 (purple) are four, three, and two times of the sensitivity of device 2 (yellow) respectively (Figure 3(c)).

## 3. Simulation details for FLP

For this model, we use the metal aluminum for accuracy. The precise geometry of the three devices shown in the inset of Figure 1(a) was traced in CleWin5 and then imported into COMSOL (Figure 4(a)). The central metal is aluminum with carrier mobility and conductivity set to 0.0012 m²/(V·s) and 7.886×10⁹ S/m . The conductivity for Al is adjusted to match the thickness of the graphene in the 2D model. Mobility and conductivity of the graphene were fitted to match the experimental data. The small deviations of the "optimal1" and "optimal2" curves (blue solid line and red solid line respectively in Figure 4(c)) from the experimental data could be due to local inhomogeneities in the EMR device, minor variations of shape or that the model is too simple; the advantage of using a simple model in this case is a reduced number of fitting parameters.

## 4. FEM and tight binding

The ballistic transport is not supported in our model of finite-element simulations. The idea of ballistic transport as an addition to the conventional EMR mechanism is an intriguing idea, as it is not clear whether an increasing ballistic contribution expected for higher quality samples or smaller-size EMR devices would lead to a higher or lower sensitivity and noise equivalent field. Such calculations could be carried out by large-scale tight-binding transport calculations or ballistic trajectory calculations using the Landauer-Büttiker framework [43]. We anticipate that large, high-quality EMR devices, where electron-phonon scattering is limiting the mobility could be used to scan different degrees of ballistic transport, by measuring the EMR characteristics from room temperature towards cryogenic temperatures, as this is known to increase the mean free path in a well-behaved manner [20].

## Author Contributions

B. Z. designed the experiment, collected and analyzed the experimental data, performed the simulation, and wrote the manuscript. K. W. and T. T. provided the hBN samples.

## Acknowledgements

B. Z. acknowledges discussions with Erik Henriksen and Peter Bøggild. B. Z. acknowledges funding from the Novo Nordisk Foundation, grant no. NNF21OC0066526 (BioMag) to financially support part of this work. K. W. and T. T. acknowledge support from the JSPS KAKENHI (Grants No. 21H05233 and No. 23H02052) and World Premier International Research Center Initiative (WPI), MEXT, Japan.

## APPENDIX A

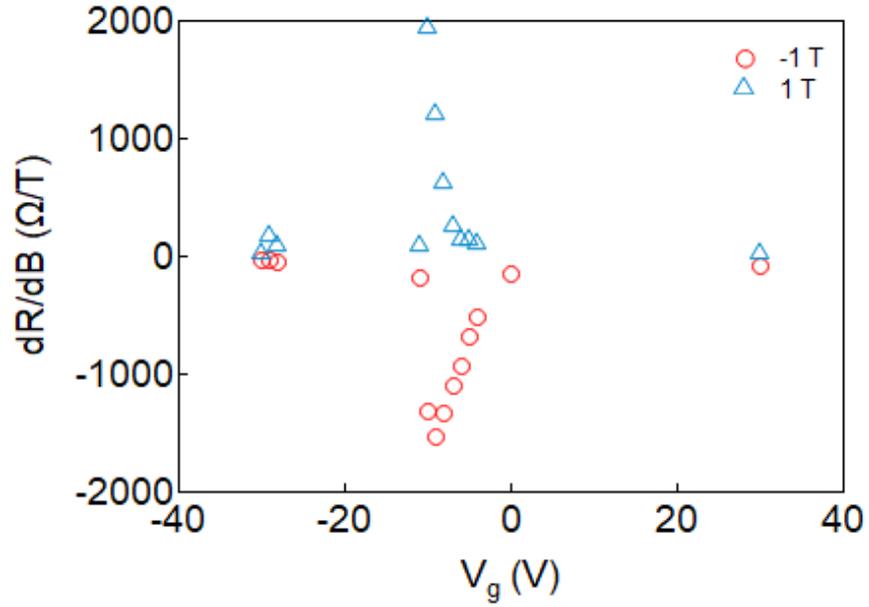

*Figure 6. Sensitivity dR/dB at +/-1T as a function of Vg, largest near CNP, corresponding to Figure 1(c).*

## APPENDIX B

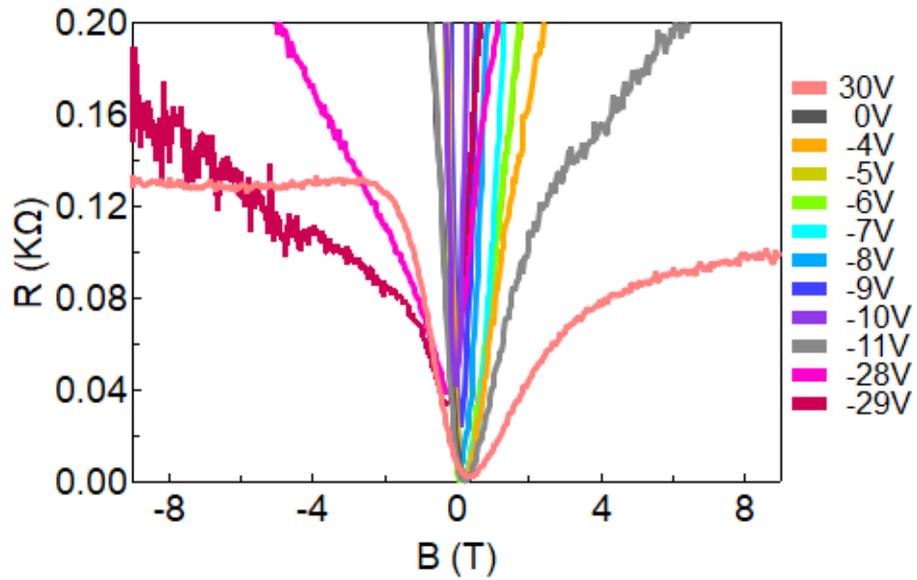

*Figure 7. Near-zero-resistance range of resistance R as a function of magnetic field B at different gate voltages Vg from 4-terminal measurements corresponding to Figure 1(b).*

## APPENDIX C

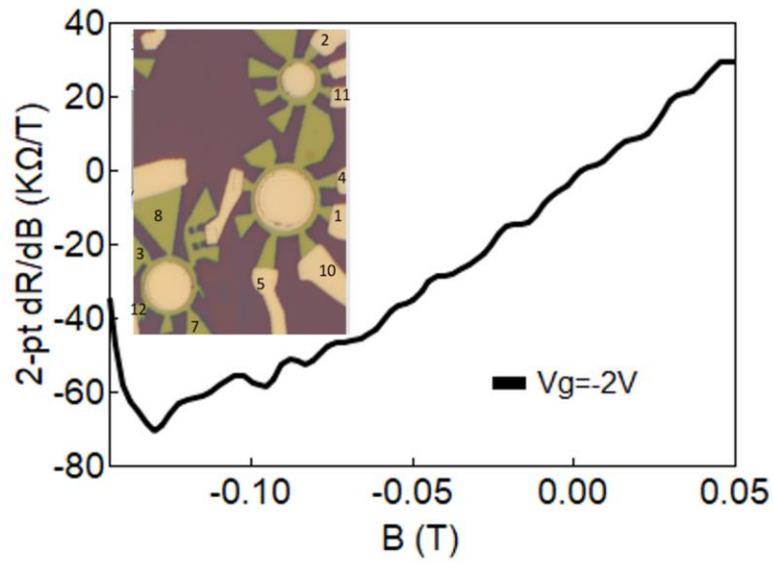

*Figure 8. The sensitivity of all three daisy chained devices as a function of B at Vg = -2 V, measured in the 2-terminal configuration at terminals 8 and 2.*

## APPENDIX D

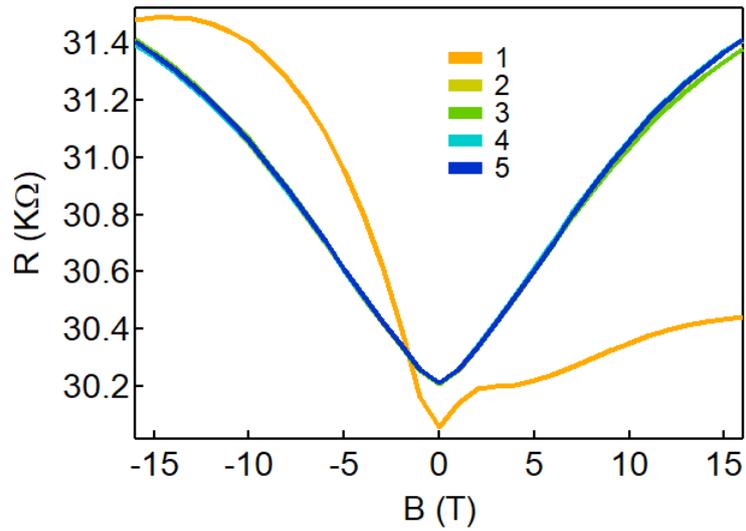

*Figure 9. The resistance of single device (1 to 5) in response to magnetic field, corresponding to Figure 3(b). The resistance of each device (2 to 5) overlaps with each other because the devices are identical. The resistance of device 1 is different from the other devices (2-5) due to the asymmetric geometry of device 1 on the end of the chain that alters the current distribution.*

## APPENDIX E

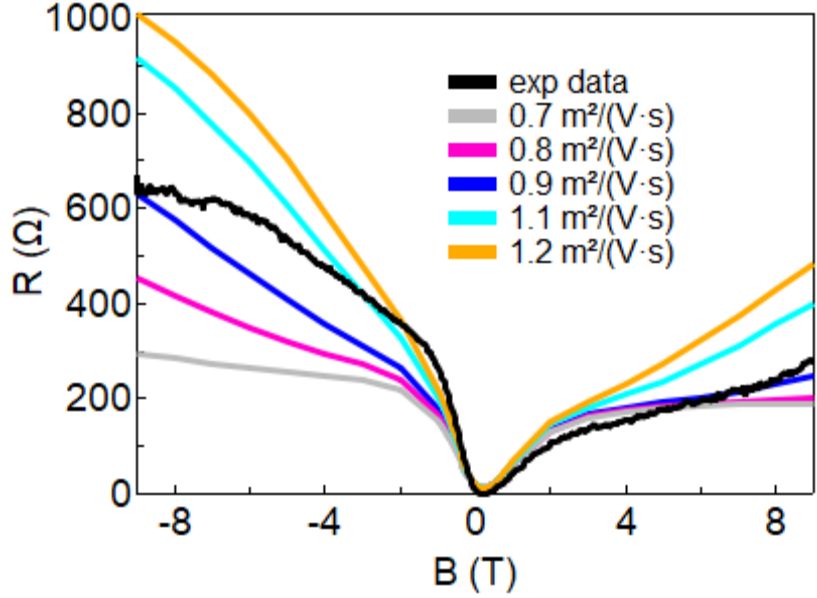

*Figure 10. Simulated resistance of graphene as a function of B. They have the same parameters as the blue solid line in Figure 4(c) except varying mobility of the outer-region graphene.*

## APPENDIX F

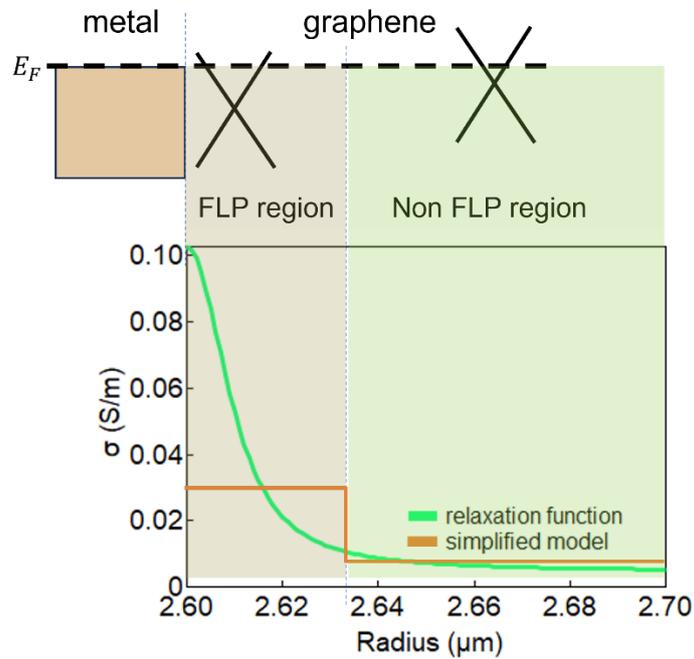

*Figure 11. Schematic of the graphene Dirac cone position relative to the Fermi energy, both near and far from a metallic contact, with the corresponding conductivities corresponding to Figure 4(b). The Dirac cone of graphene is highly doped near the metal but slightly doped far from the metal.*

## APPENDIX G

The inset in Figure 12(a) shows the three devices with same outer radius $r_0$, but different ratio $r_0/r_i$ (details in Methods). The 2-terminal measurements of one device (dashed) and three devices (solid) at gate voltages $V_g$ = -3V (black) and $V_g$ = -4V (red) are compared (Figure 12). Since changes in choice of terminals can lead to different current paths in graphene devices, two contact configurations were measured. The three-device configurations lead to largely overlapping resistance traces, and exhibit sensitivities of 29 kΩ/T and 25 kΩ/T compared to around 2 kΩ/T for the single device (Figure 12(b)). We note that such variations are not due to differences in contact resistance since the contact resistance is not expected to depend on magnetic field, and the contact resistances of terminal 5, 8, 11, 12 were measured to be in the 1-3 kΩ range.

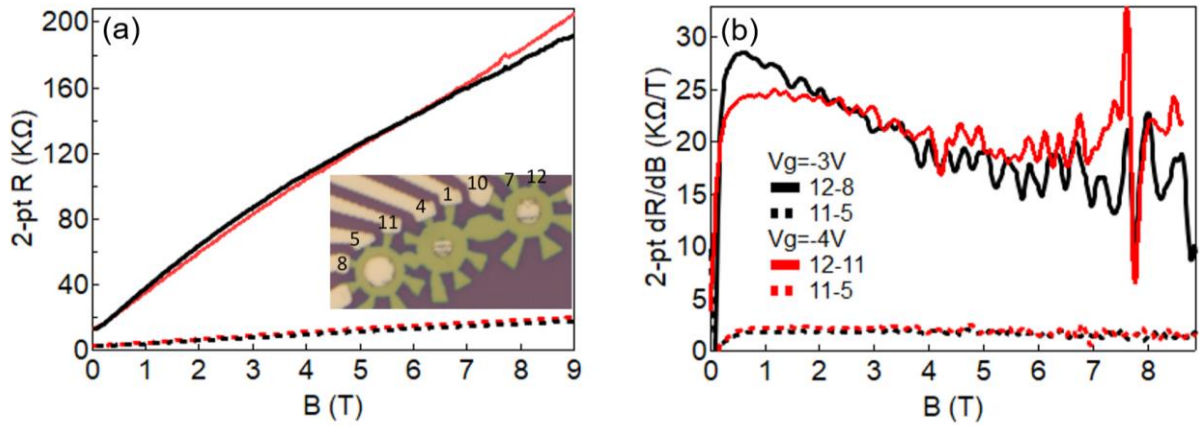

*Figure 12. (a) 2-terminal resistance and (b) corresponding sensitivity from measurements with current running through only one device (dashed, point 11-5) and three devices (solid, point 12-8 and 12-11) at gate voltages Vg = -3V (black) and -4V (red). The inset shows the optical microscope image of the three devices.*

**APPENDIX H**

From room-temperature 4-terminal measurements of the middle-sized device among the three EMR devices shown in the inset of Figure 1(a), some very strong resistance features appear at Vg = -28V, -29V and -30V far away from CNP shown in thick lines in Figure 13(a) and (c) (in log scale and linear scale, respectively). These fluctuating resistance curves rise and fall off rapidly with increasing magnetic field while all other common EMR resistance traces (thin lines) show the expected monotonical increasing trend with much smaller slopes beyond 1T. Most importantly, these oscillations reach much higher maximum resistance than all other common resistance traces: the resistance peaks from these oscillations reach around 1.9 kΩ at B = 1.2T (Vg = -29V) (black thick line), 17.3 kΩ at B = 4.4T (Vg = -30V) (red thick line), and 22 kΩ at 8.2T (Vg = -28V) (light blue thick line) and are marked by blue dashed line in Figure 13(a), while the largest common resistance trace appears at Vg = -6V (orange thin line) and saturate at only 1.7 kΩ at -9T, respectively. In linear scale, the resistance increase in these oscillations looks more prominent (Figure 13(c)). Therefore, these oscillations could enhance resistance by more than one order of magnitude.

With such a quick response to the B field, the resistance oscillations could also contribute to way higher 4-terminal sensitivities than the rest (Figure 13(b)): In contrast to the largest sensitivity 436 Ω/T reached in the common EMR traces (Figure 13(d)), the positive sensitivities of the resistance oscillations reach 2.3 kΩ/T at B= 0.45T (Vg=-29V) (black thick line), 8.5 kΩ/T at B = 2.8T (Vg = -30V) (red thick line) and 10.6 kΩ/T at B = 7.1T (Vg = -28V) (light blue thick line), while the negative sensitivities go beyond 21 kΩ/T at B = 5.2T (Vg = -30V) (red thick line) and 12 kΩ/T at B = 8.7T (Vg = -28V) (light blue thick line). We note an improvement in sensitivity of at least 50 times resulting from these resistance variations compared to the common EMR resistance response. More resistance oscillations could occur at higher Vg, which is, however, beyond the gating capability of the silicon substrate. A more systematic study could be considered to study how these oscillations arise in EMR geometry and what the maximum sensitivity it could contribute to is.

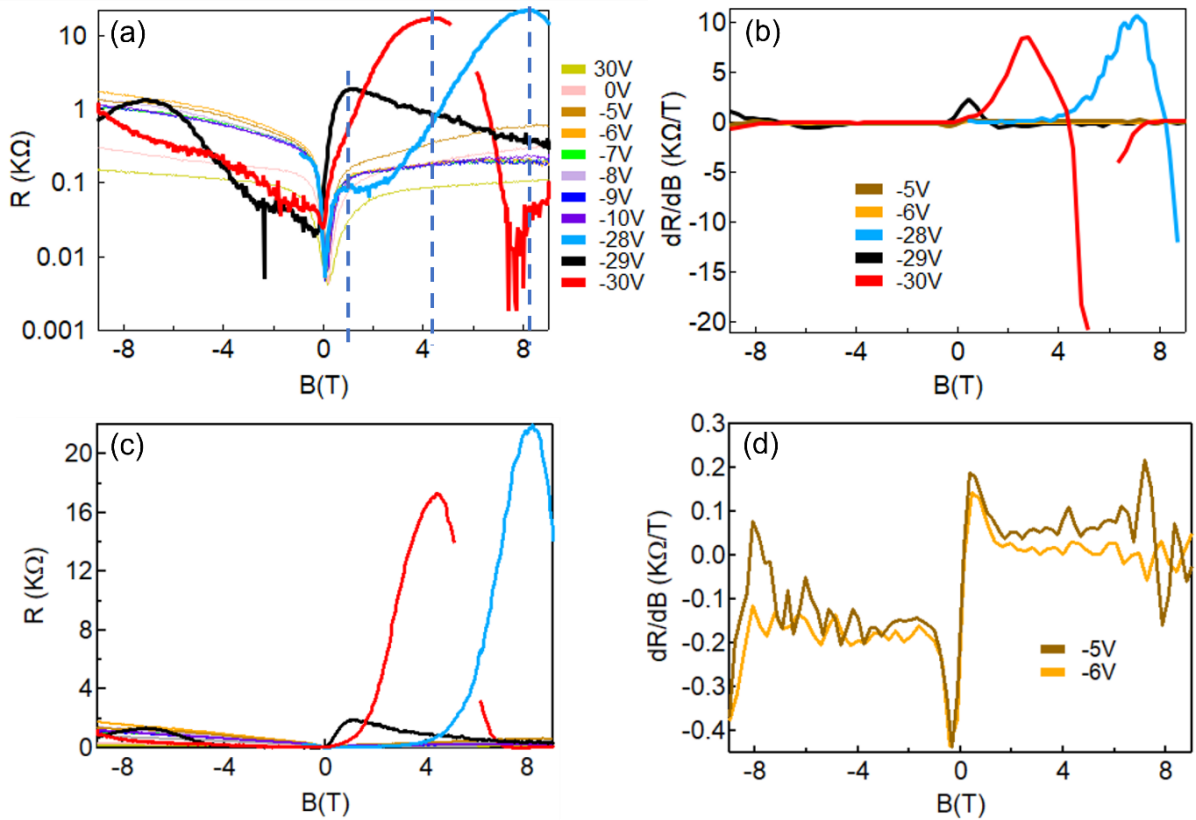

*Figure 13. (a) Strong resistance oscillations (thick lines) and common EMR resistances (thin lines) in log scale as a function of B at different Vg. (b) The 4-terminal sensitivities of three resistance oscillations and common EMR resistances as a function of B at different Vg. (c) Linear-scale version of (a). (a) and (c) share the same legend. (d) The largest two 4-terminal sensitivities among all the common EMR response as a function of B. Note that some data points in the trace at Vg = -30V (red thick line) were hidden because the signal is out of the sensitivity range of the lock-in amplifier during measurements.*